\documentclass[12pt]{article}

\usepackage{hyperref}
\usepackage{cite}
\usepackage{units}
\usepackage{epsfig}
\usepackage{pifont}
\usepackage{amsmath,amsfonts,amssymb,amsthm,mathrsfs}
\usepackage{graphicx,chngcntr,slashed}
\usepackage{cancel}
\usepackage{pstricks}
\usepackage{afterpage}
\usepackage{braket}
\usepackage{caption}
\usepackage{subcaption}
\usepackage{xcolor}

\def\mathswitch#1{\relax\ifmmode#1\else$#1$\fi}

\newcommand{\mht}{\hat{t}}
\newcommand{\mhu}{\hat{u}}
\newcommand{\mhs}{\hat{s}}

\newcommand{\mycaption}[1]{\caption{\sl #1}}

\makeatletter
\def\section{\@startsection {section}{1}{\z@}{+3.0ex plus +1ex minus
  +.2ex}{2.3ex plus .2ex}{\large\bf\boldmath}}
\def\subsection{\@startsection{subsection}{2}{\z@}{+2.5ex plus +1ex
minus +.2ex}{1.5ex plus .2ex}{\normalsize\bf\boldmath}}
\def\subsubsection{\@startsection{subsubsection}{3}{\z@}{+3.25ex plus
 +1ex minus +.2ex}{1.5ex plus .2ex}{\normalsize\it}}

\graphicspath{{.}{plots/}}

\begin{document}
\thispagestyle{empty}

\def\thefootnote{\fnsymbol{footnote}}

\begin{flushright}
\end{flushright}

\vspace{1cm}

\begin{center}

{\Large {\bf Removing flat directions in SMEFT fits: how polarized electron-ion collider data can complement the LHC}}
\\[3.5em]
{\large
Radja~Boughezal$^1$, Frank~Petriello$^{1,2}$ and Daniel~Wiegand$^{1,2}$
}

\vspace*{1cm}

{\sl
$^1$ HEP Division, Argonne National Laboratory, Argonne, Illinois 60439, USA \\[1ex]
$^2$ Department of Physics \& Astronomy, Northwestern University,\\ Evanston, Illinois 60208, USA
}

\end{center}

\vspace*{2.5cm}

\begin{abstract}

We study the potential of future Electron-Ion Collider (EIC) data to
probe four-fermion operators in the Standard Model Effective Field
Theory (SMEFT).  The ability to perform measurements with
both polarized electron and proton beams at the EIC provides a
powerful tool that can disentangle the effects from different SMEFT
operators.  We compare the potential constraints from an EIC with
those obtained from Drell-Yan data at the Large Hadron Collider.  We
show that EIC data plays an important complementary role since it probes combinations of
Wilson coefficients not accessible through available Drell-Yan
measurements.  
  
\end{abstract}

\setcounter{page}{0}
\setcounter{footnote}{0}

\newpage


\section{Introduction}

\noindent
The Standard Model (SM) of particle physics has so far been successful in
describing all observed laboratory phenomena.  No new particles beyond those
present in the SM have been discovered at the Large Hadron Collider
(LHC) or in other experiments, and no appreciable deviation
from SM predictions has been conclusively observed.  Given this
situation it is increasingly important to understand how indirect
signatures of new physics can be probed and constrained by the
available data.  This effort will help guide future searches for
new physics by suggesting in what channels measurable
deviations from SM predictions may occur given the current bounds.

A convenient theoretical framework for investigating indirect
signatures of heavy new physics without associated new particles is the SM
effective field theory (SMEFT) containing higher-dimensional operators
formed from SM fields.  The leading dimension-6 operator basis of
SMEFT for on-shell fields has been completely classified~\cite{Buchmuller:1985jz, Grzadkowski:2010es} (there is a
dimension-5 operator that violates lepton number which we do not
consider here).  Considerable effort has been devoted to performing global
analyses of the availabe data within the SMEFT
framework~\cite{Han:2004az,Pomarol:2013zra,Chen:2013kfa,Ellis:2014dva,Wells:2014pga,Falkowski:2014tna,deBlas:2016ojx,Cirigliano:2016nyn,Biekotter:2018rhp,Grojean:2018dqj,Hartland:2019bjb,Brivio:2019ius}.
There are numerous questions that must be addressed when performing
global fits within the dimension-6 SMEFT framework, including the need
for higher-order corrections in the SM coupling constants~\cite{Passarino:2016pzb}, the
importance of effects from dimension-8 and beyond~\cite{Henning:2015alf,Degrande:2013kka,Azatov:2016sqh,Hays:2018zze,Alioli:2020kez
}, and the estimation of theoretical errors~\cite{Keilmann:2019cbp}.

Another issue that arises in global fits to the SMEFT parameter space is the appearance of
flat directions that occur when the available experimental
measurements cannot disentangle the contributions from different Wilson
coefficients.  These flat directions may be either
exact or approximate.  There are many examples of this
phenomenon.  For example, it is well known that Higgs cross section
measurements alone cannot distinguish between new-physics corrections
to the Higgs couplings to gluons and top
quarks~\cite{Azatov:2013xha}.  Our focus here will be on 2-lepton,
2-quark four-fermion operators appearing in the SMEFT.  The presence
of operator combinations not probed by the available low-energy data has
been discussed in the literature~\cite{Falkowski:2017pss}.  The
expectation is that these operators are well-probed by high
invariant-mass Drell-Yan distributions at the LHC, which has both
large integrated luminosity and the requisite high energy for which
we expect potential SMEFT corrections to become important.  There have indeed
been numerous studies of the importance of Drell-Yan measurements in
constraining four-fermion
operators~\cite{Falkowski:2017pss,Dawson:2019xfp}.  However, only a few combinations of Wilson coefficients can be probed
in principle by Drell-Yan measurements, a point made previously in the
literature~\cite{Alte:2018xgc}.  In practice only a subset of even
these combinations can be probed due to the nature of the current
experimental studies, as we discuss later.  Future
analyses of constraints on SMEFT operators will need to identify new
data sets that measure the combinations not determined by Drell-Yan
production at the LHC, such as flavor observables where one-loop
effects can help break degeneracies~\cite{Hurth:2019ula, Aoude:2020dwv}.

Our goal in this manuscript is to illustrate the important role that
future polarized deep-inelastic scattering (DIS) experiments may play in
the study of SMEFT, and in particular in disentangling the effects of four-fermion operators indistinguishable
at the LHC.  In the coming decade the construction of an electron-ion
collider (EIC) with polarization of both electron and proton beams is
expected, and high-precision polarized electron-proton data will
become available.  Some studies of new physics searches possible
with an EIC have been performed~\cite{Accardi:2012qut}.  However, we
are aware of no detailed investigation of
what aspects of the SMEFT may be probed at an EIC, and in particular its usefulness
in studying combinations of Wilson coefficients not accessible at the LHC.
In this paper we consider the following points.
\begin{itemize}

\item We study the deviations induced by dimension-6 four-fermion
  operators in the SMEFT on both polarized and unpolarized
  charged-current and neutral-current DIS.  We
  determine which regions of parameter space are sensitive to 
  dimension-6 Wilson coefficients.  We show that the deviations
  allowed by current constraints are larger than the current parton
  distribution function (PDF) errors for both polarized and unpolarized
  protons.  Since the SM DIS hard-scattering cross sections are known
  through next-to-next-to-leading order in QCD~\cite{Zijlstra:1992qd,Zijlstra:1993sh}, theoretical errors
  should not be a limiting factor in studies of the SMEFT at an EIC.

 \item We review the contributions of dimension-6 SMEFT operators to
   neutral-current Drell-Yan at the LHC, and analytically demonstrate
   the appearance of approximate flat directions in the space of
   Wilson coefficients.  We show the current experimental measurements
   at the LHC
   are not well-suited to SMEFT studies.  High invariant-mass
   forward-backward asymmetry measurements would allow additional
   probes of the SMEFT parameter space, a point also emphasized in
   Ref.~\cite{Alte:2018xgc}.  However, even with such observables
   many combinations of Wilson coefficients remain poorly tested in LHC
   Drell-Yan production and would benefit from polarized DIS measurements.

  \item We perform fits to Drell-Yan data from the LHC to
    numerically illustrate the flat directions.  We identify several
    example choices of Wilson coefficients that demonstrate the types
    of degeneracies that appear at the LHC.  We show how data from a future EIC is
    complementary to that obtained from the LHC and can better probe
    certin combinations of Wilson coefficients. Combined fits of LHC
    and projected EIC data lead to much stronger constraints than
    either experiment alone.  We show that the ability to polarize both
    electron and proton beams at an EIC is crucial in obtaining these
    projected bounds.

\end{itemize}  

Our paper is organized as follows.  We review the aspects of the
four-fermion operators in the SMEFT relevant to our analysis in
Section~\ref{sec:smeft}.  In Section~\ref{sec:dis} we present the
formulae needed for the study of unpolarized and polarized DIS.
We study the phenomenology of SMEFT contributions to DIS at an EIC in
Section~\ref{sec:eic}.  We study neutral-current Drell-Yan production
of lepton pairs at the LHC in Section~\ref{sec:DY}, where we also
demonstrate the appearance of flat directions in the space of Wilson
coefficients.  In Section~\ref{sec:fits} we present the main results of
our paper, fits to the LHC and projected EIC data for a range of
different scenarios.  We emphasize the complementarity of the two
experiments, and show the importance of polarized measurements at the
EIC.  Finally, we conclude in Section~\ref{sec:conc}.

\section{Review of the SMEFT} \label{sec:smeft}

We review in this section aspects of the SMEFT relevant for
our analysis of DIS and Drell-Yan.  The SMEFT is an extension of the SM Lagrangian to include terms
suppressed by an energy scale $\Lambda$ at which the ultraviolet completion
becomes important and new particles beyond the SM appear.  Truncating the expansion in $1/\Lambda$ at dimension-6, and
ignoring operators of odd-dimension which violate lepton number, we
have
\begin{equation}
{\cal L} = {\cal L}_{SM}+ \sum_i C_{i} {\cal
  O}_{i} + \ldots,
\end{equation}
where the ellipsis denotes operators of higher dimensions.  The Wilson
coefficients defined above have dimensions of $1/\Lambda^2$.  When computing cross sections we consider only the leading
interference of the SM amplitude with the dimension-6 contribution.
This is consistent with our truncation of the SMEFT expansion above,
since the dimension-6 squared contributions are formally the
same order in the $1/\Lambda$ expansion as the dimension-8 terms which we neglect.
The
following four-fermion operators in Table~\ref{tab:ffops} can affect both DIS and Drell-Yan
at leading-order in the coupling constants for massless fermions,
which we assume here.
\begin{table}[h!]
\centering
\begin{tabular}{|c|c||c|c|}
\hline
${\cal O}_{lq}^{(1)}$ & $(\bar{l}\gamma^{\mu} l) (\bar{q}\gamma_{\mu} q)$ &  ${\cal O}_{lu}$ & $(\bar{l}\gamma^{\mu} l) (\bar{u}\gamma_{\mu}  u)$ 
  \\
  ${\cal O}_{lq}^{(3)}$ & $(\bar{l}\gamma^{\mu} \tau^I l)
                     (\bar{q}\gamma_{\mu} \tau^I l q)$ & ${\cal O}_{ld}$ & $(\bar{l}\gamma^{\mu} l) (\bar{d}\gamma_{\mu}  d)$   
  \\
  ${\cal O}_{eu}$ & $(\bar{e}\gamma^{\mu} e) (\bar{u}\gamma_{\mu}  u)$
                                                                     & ${\cal O}_{qe}$ & $(\bar{q}\gamma^{\mu} q) (\bar{e}\gamma_{\mu}  e)$             
  \\
  ${\cal O}_{ed}$ & $(\bar{e}\gamma^{\mu} e) (\bar{d}\gamma_{\mu}  d)$
                                                                     & &
  \\
  \hline
\end{tabular}
\mycaption{Dimension-6 four-fermion operators contributing to DIS and
  DY at leading order in the coupling constants.\label{tab:ffops}}
\end{table}
$q$ and $l$ denote left-handed quark and lepton doublets, while
$u$, $d$ and $e$ denote right-handed singlets for the up quarks, down
quarks and leptons, respectively.  $\tau^I$ denote the SU(2)
Pauli matrices.  We have suppressed flavor indices
for these operators, and in our analysis we assume flavor universality
for simplicity.  We note that the overall
electroweak couplings that govern lepton-pair production are also
shifted in the SMEFT by operators other than those considered above.  Such contributions are far better bounded
through other data sets such as precision $Z$-pole
observables~\cite{Dawson:2019xfp}, and we neglect them here.  The
above assumptions leave us with the seven Wilson coefficients
associated with the operators in Table~\ref{tab:ffops} entering
the predictions for our cross sections.

\section{Review of DIS formalism}\label{sec:dis}

We review in this section the relevant formulae describing both
unpolarized and polarized DIS in the process $l(k) +P(P) \to
l^{\prime}(k^{\prime})+X$, where $P$ denotes a proton.  We
consider the leading-order partonic process $l(k)+q(p) \to
l^{\prime}(k^{\prime}) +q_f(p_f)$, including both the SM contributions and
the corrections induced by dimension-6 SMEFT operators.   Expressions
for the charged-current process are also given below. The relation between partonic and hadronic momenta is $p=xP$.  It is standard to introduce the
momentum transfer $q = k-k^{\prime}$, with $q^2 = -Q^2$.  We recall here some of the
basic kinematic relations relevant for DIS:
\begin{equation}
 p\cdot k = \frac{xs}{2},\;\; p_f \cdot k^{\prime} = \frac{xs}{2}, \;\;
  k \cdot k^{\prime} = \frac{Q^2}{2}, \;\; P \cdot q = \frac{Q^2}{2x}, \;\;\; p \cdot q = \frac{Q^2}{2}, \;\; \frac{P \cdot q}{P \cdot k} = \frac{p \cdot q}{p \cdot k} =y.
  \label{eq:variables}
\end{equation}
We can use these relations to show that $Q^2 = xys$ at leading-order.

The matrix elements receive SM contributions from both photon and
$Z$-boson exchange.  In the SMEFT there is an additional correction from
four-fermion contact interactions.  We can split the differential
cross section into the following contributions that arise from the
interference of the relevant diagrams:
\begin{equation}
\frac{d^2\sigma}{dx \, dQ^2} = \frac{4 \pi \alpha^2}{x Q^4} \sum_q 
f_{q,\lambda_q}(x,Q^2) \left\{\frac{d^2\sigma^{\gamma\gamma}}{dx dQ^2}
  +\frac{d^2\sigma^{\gamma Z}}{dx dQ^2}
  +\frac{d^2\sigma^{Z Z}}{dx dQ^2}
  +\frac{d^2\sigma^{\gamma SMEFT}}{dx dQ^2}
  +\frac{d^2\sigma^{Z SMEFT}}{dx dQ^2}
\right\}.
\label{eq:DIScr}
\end{equation}  
We have used $\lambda_e$ and $\lambda_q$ to respectively denote the helicities of
the lepton and quark that enter the hard-scattering process.  For
fully-polarized states, $\lambda_i = \pm 1$ in our normalization.  The leading-order expressions for the SM contributions are given
below:
\begin{eqnarray}
  \frac{d^2\sigma^{\gamma\gamma}}{dx dQ^2} &=& x Q_q^2 \left[
                                               (1-y)+\frac{y^2}{2}+\frac{\lambda_q\lambda_e}{2}y(2-y)
                                               \right], \nonumber \\
  \frac{d^2\sigma^{\gamma Z}}{dx dQ^2} &=& x \frac{e_q N_{\gamma Z}}{2}
                                           \left[ g_L^q
                                           g_L^e(1-\lambda_q)(1-\lambda_e)+g_R^q
                                           g_R^e(1+\lambda_q)(1+\lambda_e) \right.
                                           \nonumber \\
                                           && \left. +g_R^qg_L^e(1-y)^2(1+\lambda_q)(1-\lambda_e)
                                           +g_L^qg_R^e(1-y)^2(1-\lambda_q)(1+\lambda_e)
                                              \right], \nonumber \\
  \frac{d^2\sigma^{ZZ}}{dx dQ^2} &=&x \frac{N_{ZZ}}{4}
                                           \left[ (g_L^q
                                           g_L^e)^2(1-\lambda_q)(1-\lambda_e)+(g_R^q
                                           g_R^e)^2(1+\lambda_q)(1+\lambda_e) \right.
                                           \nonumber \\
                                           && \left. +(g_R^qg_L^e)^2(1-y)^2(1+\lambda_q)(1-\lambda_e)
                                           +(g_L^qg_R^e)^2(1-y)^2(1-\lambda_q)(1+\lambda_e)
                                              \right].
\label{eq:SMDIS}                                             
\end{eqnarray}
We have introduced the following abbreviations in these expressions:
\begin{equation}
N_{\gamma Z} = \frac{G_F M_Z^2}{2\sqrt{2} \pi \alpha}
\frac{Q^2}{Q^2+M_Z^2},\;\;\; N_{ZZ} = N_{\gamma Z}^2.
\end{equation}  
For the
SM left-handed and right-handed fermion couplings we follow the
conventions of Ref.~\cite{Denner:1991kt}:
\begin{equation}
g_L^f = I_3^f-Q_f s_W^2, \;\;\; g_R^f = -Q_f s_W^2.
\end{equation}  

We give below the expressions for the SMEFT corrections in the
up-quark initial state:
\begin{eqnarray}
  \frac{d^2\sigma_u^{\gamma SMEFT}}{dx dQ^2} &=& -x \frac{Q_u
                                               Q^2}{8\pi\alpha} \left[
                                               C_{eu}(1+\lambda_u)(1+\lambda_e)
                                               +(C_{lq}^{(1)}-C_{lq}^{(3)})
                                               (1-\lambda_u)(1-\lambda_e)
                                               \right.
                                               \nonumber \\
                                               & & \left. +(1-y)^2 C_{lu}
                                               (1+\lambda_u)(1-\lambda_e)
                                               +(1-y)^2 C_{qe}
                                               (1-\lambda_u)(1+\lambda_e) \right]
                                             \nonumber \\
  \frac{d^2\sigma_u^{Z SMEFT}}{dx dQ^2} &=&-x \frac{N_{\gamma Z}
                                               Q^2}{8\pi\alpha} \left[
                                              g_R^u g_R^e C_{eu}(1+\lambda_u)(1+\lambda_e)
                                               + g_L^u g_L^e (C_{lq}^{(1)}-C_{lq}^{(3)})
                                               (1-\lambda_u)(1+\lambda_e)
                                               \right.
                                               \nonumber \\
                                               & & \left. + g_R^u g_L^e (1-y)^2 C_{lu}
                                               (1+\lambda_u)(1-\lambda_e)
                                               + g_L^u g_R^e (1-y)^2 C_{qe}
                                               (1-\lambda_u)(1+\lambda_e)
                                                   \right].
\label{eq:DISSMEFT}                                                   
\end{eqnarray}  
To obtain results for the down-quark initial state, we simply make the
following replacements in the formulae above:
\begin{equation}
Q_u \to Q_d, \;\; g_{L,R}^u \to g_{L,R}^d,\;\; C_{lu} \to C_{ld}, \;\;
C_{eu} \to C_{ed}, \;\; C_{lq}^{(3)} \to -C_{lq}^{(3)}.
\end{equation}  

From these formulae we can obtain the results for the polarized and
unpolarized cross sections.  The unpolarized cross section is obtained
by averaging over the two quark helicity possibilities $\lambda_q= \pm
1$ and setting the PDF in Eq.~(\ref{eq:DIScr}) to the usual
unpolarized one, while the polarized result is obtained by taking the difference 
$\lambda_q=-1$ minus $\lambda_q=+1$ and interpreting the PDF in
Eq.~(\ref{eq:DIScr}) as the usual polarized PDF.  Upon forming these
two combinations we obtain four physically observable differential
cross sections in neutral-current DIS: the polarized and unpolarized
cross sections with positive or negative $\lambda_e$.

We briefly present here the formulae for the charged-current process
$\nu_{\mu}(k)+u(p) \to \mu(k^{\prime})+d(p_f)$. We directly show the
results for the unpolarized and polarized partonic cross sections.  The SM differential cross sections are
\begin{eqnarray}
\frac{d \sigma^{WW}_{unpol} }{dxdQ^2} &=& \frac{g^4
  (1-\lambda_e)}{64 \pi (Q^2+M_W^2)^2}, \nonumber \\
\frac{d \Delta \sigma^{WW}}{dxdQ^2} &=& \frac{g^4
  (1-\lambda_e)}{32 \pi (Q^2+M_W^2)^2}.
\end{eqnarray}
The corrections coming from SMEFT four-fermion operators are
\begin{eqnarray}
\frac{d \sigma^{WSMEFT}_{unpol} }{dxdQ^2} &=& -\frac{g^2
  (1-\lambda_e) C_{lq}^{(3)}}{8 \pi (Q^2+M_W^2)}, \nonumber \\
\frac{d \Delta \sigma^{WSMEFT}}{dxdQ^2} &=& -\frac{g^2
  (1-\lambda_e) C_{lq}^{(3)}}{4 \pi (Q^2+M_W^2)}.
\end{eqnarray}
We note that only the left-handed polarization state contributes.  

\section{Phenomenology of DIS at the EIC} \label{sec:eic}

In this section we briefly review the expected parameters of an EIC,
and study the deviations induced by the four-fermion SMEFT operators
considered above on both neutral and charged-current DIS.  The recently announced EIC at Brookhaven
National Laboratory will be a high-energy and
high-luminosity tool to investigate the structure of
nucleons and nuclei.  The physics potential of the EIC is detailed in
Ref.~\cite{Accardi:2012qut}, as are the various machine parameters
assumed below in our study.  It is planned to be tunable over a large range of energies,
different polarizations and types of heavy ions, as well as protons.  The machine is projected to operate at a center-of-mass-energy
approaching $\sqrt{s} \approx 140\;\textrm{GeV}$, which we assume in
our study.  We assume that it will collect $10\,\textrm{fb}^{-1}$, which
we split equally among the four modes identified in the previous
section (polarized and unpolarized with both positive and negative $\lambda_e$).  We also
study the impact of accumulating $100\,\textrm{fb}^{-1}$.  We
assume that the EIC will reach 70\% polarization for both proton and
electron beams.


\subsection{Standard Model Contributions}

\begin{figure}[h!]
  \centering
    \includegraphics[width=6.5in]{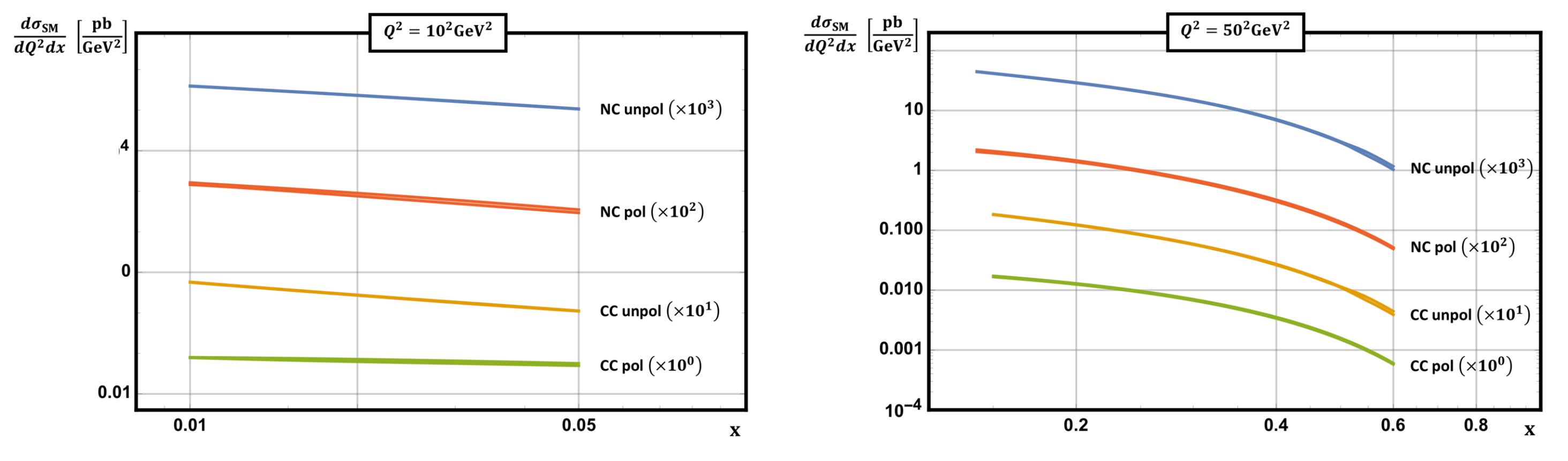}
    \caption{Comparison of unpolarized and polarized Standard Model
      cross sections for neutral and charged current processes for
      different values of $Q^2$.  The cross sections assume $\lambda_e = -0.7$. The $1-\sigma$ error band stems from the uncertainty of the corresponding PDFs.} 
  \label{fig:SMDIS}
\end{figure}

We begin by briefly summarizing and discussing the Standard Model
predictions for the different cross sections that will be measured at
the future EIC.  The expressions in Eq.~(\ref{eq:SMDIS}) are evaluated with the electroweak input parameters~\cite{Tanabashi:2018oca}: 
\begin{align}
\alpha^{-1} &= 137.036, \;\;\; G_F = 1.16638 \times 10^{-5} \,\textrm{GeV}^{-2},  \nonumber\\
M_Z &= 91.1876\,\textrm{GeV}, \;\;\; M_W = 80.379\,\textrm{GeV}.
\end{align}
We use the {\sc NNPDF3.1 NLO}~\cite{Ball:2017nwa} PDFs in
the unpolarized case and {\sc NNPDFPol1.1}~\cite{Nocera:2014gqa} in
the polarized case throughout.  We constrain the angular variable $y$,
defined in Eq.~(\ref{eq:variables}), to be between $0.1$ and $0.9$ for
both the neutral and charged current processes, in accordance with
values quoted in the literature~\cite{Chu:2017mnm,Aschenauer:2013iia}. To avoid non-perturbative QCD effects impacting
our analysis we only consider values of $Q^2$ above
$(12\;\textrm{GeV})^2$. The momentum fraction $x$ is constrained in our fits to be below $0.2$.
To provide some intuition regarding the expected evant rates at the
EIC we show the SM cross sections for both charged and neutral current processes in Fig.~\ref{fig:SMDIS}.

\subsection{SMEFT Contributions}
\begin{figure}[h!]
  \centering
    \includegraphics[width=6.5in]{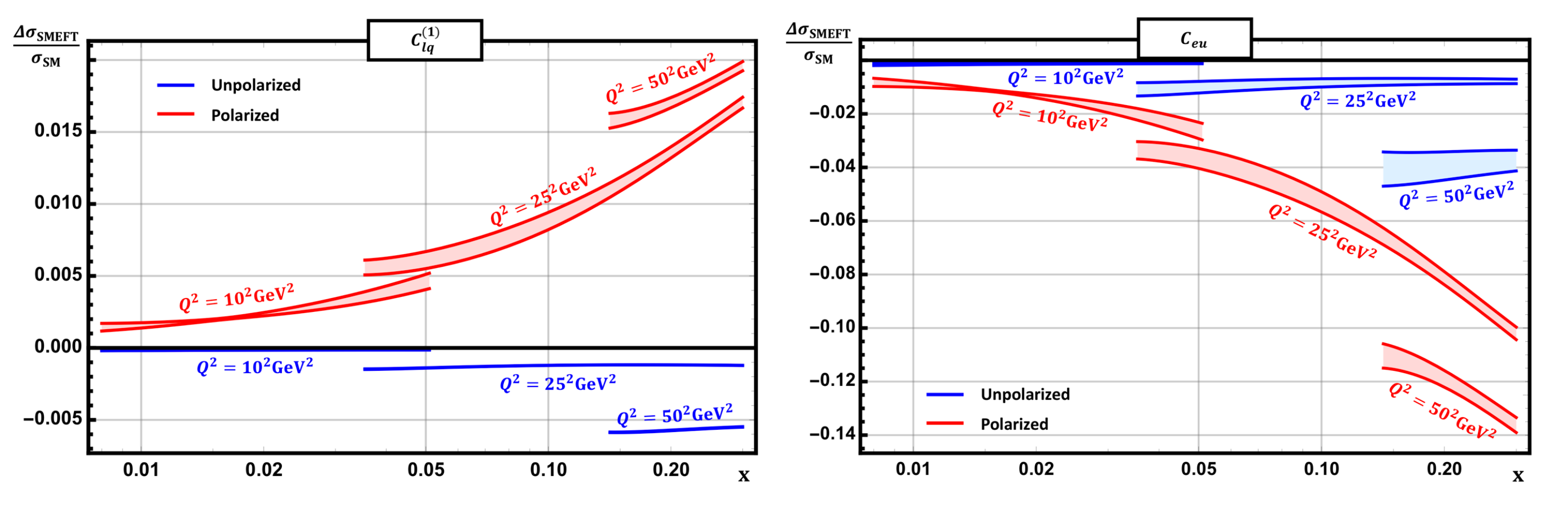}
    \caption{Neutral-current SMEFT deviation normalized to Standard
      Model predictions for the Wilson coefficients
      $C_{lq}^{(1)}$ and $C_{eu}$ as a
      function of Bjorken-$x$ for different choices of $Q^2$. The error
      bands illustrate the the $1-\sigma$ interval stemming from the
      uncertainty of the PDFs.  The SMEFT deviations for the
      unpolarized cross section are in blue, and for the polarized
      cross section in red.  We note that the kinematic constraint on
    $y$ leads to the turn-on of the curves at different $x$-values for
  each $Q^2$ choice.}
  \label{fig:ratioSMEFT}
\end{figure}

We next allow for SMEFT four-fermion operator contributions to modify our
observables. We assume $C_i =  1/\textrm{TeV}^2$ in order to illustrate these effects.  First we illustrate the potential of the DIS
observables by plotting the expected relative deviation from the
Standard Model over a large part of $(x,Q^2)$ space in
Fig.~\ref{fig:ratioSMEFT} for two example Wilson coefficients.  We see that the deviations grow with both $x$ and $Q^2$, indicating
that these phase-space regions will be most sensitive to the SMEFT
effects.  It is evident from these plots that PDF uncertainties are sub-dominant
to potential SMEFT deviations, even in the case of a polarized proton
beam.  We
also note that the expected deviations become large relative to the
expected precision of the EIC.  

\begin{figure}[h!]
  \centering
    \includegraphics[width=6.5in]{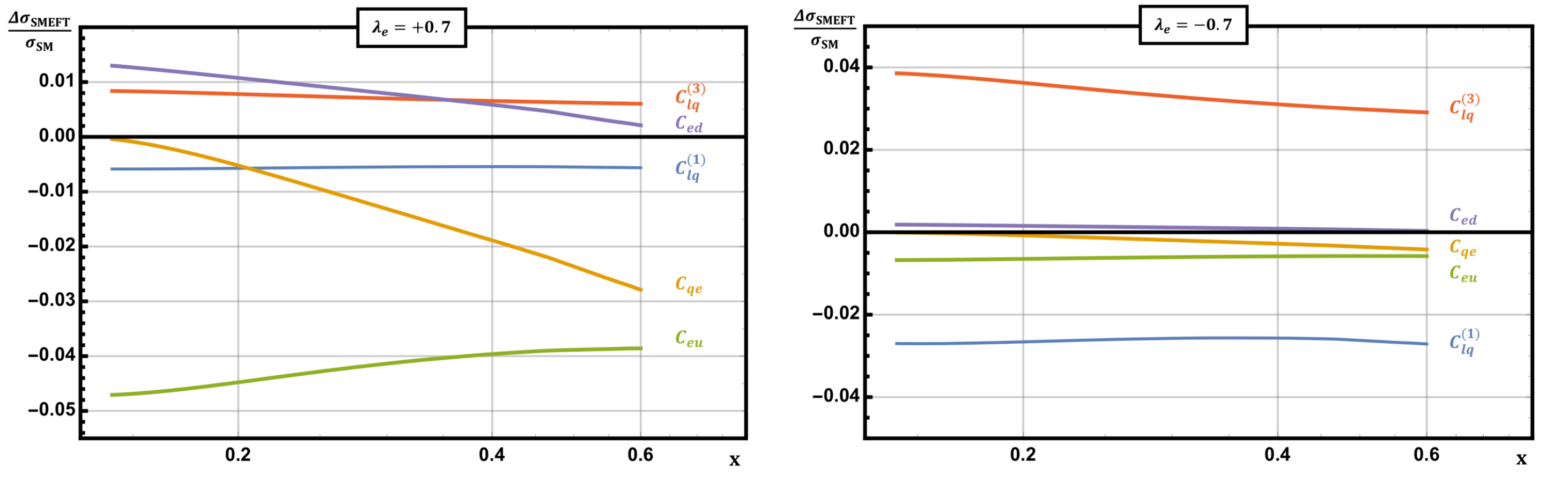}
    \caption{Comparison of the size of the unpolarized SMEFT
      deviations for each of the Wilson coefficients
      $C_{qe}$, $C_{eu}$, $C_{ed}$, $C_{lq}^{(1)}$ and $C_{lq}^{(3)}$ for
      different values of the electron polarization.  The plots are
      for fixed $Q^2 = (50\, \textrm{GeV})^2$.}
  \label{fig:SMEFTcomparison}
\end{figure}

We now show how different observables are sensitive to different
combinations of Wilson coefficients. This is illustrated in
Fig.~\ref{fig:SMEFTcomparison}, where we compare the relative
deviations for each of the Wilson coefficients
switched on separately for different electron polarizations. We see
that for positive electron polarization we primarily probe $C_{qe}$ and
$C_{eu}$, while $C_{lq}^{(1)}$ and $C_{lq}^{(3)}$ only lead to a small
shift of the cross section. For negative electron polarization we find the opposite
behavior.  We will see later that this ability of the EIC to
discriminate between different Wilson coefficients using polarized
observables can help probe SMEFT effects difficult to see at the LHC.
%

\section{Neutral-current Drell-Yan in the SMEFT} \label{sec:DY}

We discuss here the Drell-Yan process at the LHC.  Our analysis is
performed at leading order in the SMEFT.  A partial calculation of the higher-order terms is given in Ref.~\cite{Dawson:2019xfp}.  Our major focus
will be on identifying combinations of Wilson coefficients for which
the SMEFT-induced corrections vanish.  This shows that the LHC is not
sensitive to these combinations, making potential EIC probes
important.  We will define four example
choices of Wilson coefficients that allow us to compare the
sensitivities of the LHC and a future EIC in different scenarios.

\subsection{Review of Drell-Yan formulae}

We first present formulae for the partonic channel $u(p_1)\bar{u}(p_2) \to l(p_3)
\bar{l}(p_4)$. Three diagrams contribute to this process: photon exchange, $Z$-boson
exchange, and a four-fermion contact interaction.  It is
straightforward to derive the differential cross section for this
process.  We split it into the following contributions, labeled by
which diagrammatic interference they arise from:
\begin{eqnarray}
\frac{d\sigma_{u\bar{u}}}{dM^2 dY dc_{\theta}} &=& \frac{1}{32 \pi M^2
  s} f_u(x_1) f_{\bar{u}}(x_2) \left\{
  \frac{d\hat{\sigma}^{\gamma\gamma}_{u\bar{u}}}{dM^2 dY dc_{\theta}}
  +  \frac{d\hat{\sigma}^{\gamma Z}_{u\bar{u}}}{dM^2 dY dc_{\theta}}
  +  \frac{d\hat{\sigma}^{ZZ}_{u\bar{u}}}{dM^2 dY dc_{\theta}}
  \right. \nonumber \\
 &+& \left.
    \frac{d\hat{\sigma}^{\gamma SMEFT}_{u\bar{u}}}{dM^2 dY dc_{\theta}}
  +  \frac{d\hat{\sigma}^{Z SMEFT}_{u\bar{u}}}{dM^2 dY dc_{\theta}}
\right\}.
\end{eqnarray}  
Here, $x_1$ and $x_2$ are the Bjorken momentum fractions of the
partons from each proton, $M^2$ and $Y$ are respectively the invariant
mass and rapidity of the di-lepton system, and $c_{\theta}$ is the
cosine of the CM-frame scattering angle of the negatively-charged
lepton.  To obtain the full hadronic cross section from this partonic
channel we integrate this over $x_1$ and $x_2$.  The separate contributions from each diagrammatic
interference are given below:
\begin{eqnarray}
 \frac{d\hat{\sigma}^{\gamma\gamma}_{u\bar{u}}}{dM^2 dY dc_{\theta}}&=& \frac{32 \pi^2 \alpha^2
                                Q_u^2}{3}\frac{\mht^2+\mhu^2}{\mhs^2},\nonumber
  \\
\frac{d\hat{\sigma}^{\gamma Z}_{u\bar{u}}}{dM^2 dY dc_{\theta}} &=& -\frac{8 \pi \alpha Q_u g_Z^2
                                }{3}\frac{(g_R^u
                            g_L^e+g_R^eg_L^u)\mht^2+(g_R^u
                            g_R^e+g_L^eg_L^u)\mhu^2}{\mhs (\mhs-M_Z^2)},
                            \nonumber \\
\frac{d\hat{\sigma}^{ZZ}_{u\bar{u}}}{dM^2 dY dc_{\theta}} &=& \frac{ g_Z^4
                                }{3}\frac{((g_R^u
                            g_L^e)^2+(g_R^eg_L^u)^2)\mht^2+((g_R^u
                            g_R^e)^2+(g_L^eg_L^u)^2)\mhu^2}{(\mhs-M_Z^2)^2},
                        \nonumber \\
  \frac{d\hat{\sigma}^{\gamma SMEFT}_{u\bar{u}}}{dM^2 dY dc_{\theta}}&=& -\frac{8 \pi \alpha Q_u 
                                }{3}\frac{(C_{lu}+C_{qe})\mht^2+(C_{eu}+C_{lq}^{(1)}-C_{lq}^{(3)})\mhu^2}{\mhs },
                            \nonumber \\
 \frac{d\hat{\sigma}^{Z SMEFT}_{u\bar{u}}}{dM^2 dY dc_{\theta}}&=& \frac{2g_Z^2
                                }{3}\frac{(g_R^ug_L^eC_{lu}+g_R^eg_L^uC_{qe})\mht^2+(g_R^ug_R^eC_{eu}+g_L^ug_L^eC_{lq}^{(1)}-g_L^ug_L^eC_{lq}^{(3)})\mhu^2}{\mhs-M_Z^2}.
\label{eq:DYMEs}                           
\end{eqnarray}
We have identified the usual partonic Mandelstam invariants
$\mhs=(p_1+p_2)^2$, $\mht=(p_1-p_3)^2$, $\mhu=(p_1-p_4)^2$.  They
depend upon the scattering angle $c_{\theta}$ according to
\begin{equation}
\mht = -\frac{\mhs}{2}(1-c_{\theta}),\;\;\; \mhu = -\frac{\mhs}{2}(1+c_{\theta}).
\end{equation}  
For the
SM left-handed and right-handed fermion couplings we follow the
conventions of Ref.~\cite{Denner:1991kt}:
\begin{equation}
g_L^f = I_3^f-Q_f s_W^2, \;\;\; g_R^f = -Q_f s_W^2.
\end{equation}  
We note that we can obtain the
partonic channel $\bar{u}(p_1)u(p_2) \to  l(p_3)
\bar{l}(p_4)$ by interchanging $ \mht \leftrightarrow \mhu$.  To obtain results for the down-quark initiated process
$d(p_1)\bar{d}(p_2) \to  l(p_3) \bar{l}(p_4)$ we make the following
changes in Eq.~(\ref{eq:DYMEs}):
\begin{equation}
Q_u \to Q_d, \;\; g_{L,R}^u \to g_{L,R}^d,\;\; C_{lu} \to C_{ld}, \;\;
C_{eu} \to C_{ed}, \;\; C_{lq}^{(3)} \to -C_{lq}^{(3)}.
\label{eq:uptodown}
\end{equation}  
The sign change for $C_{lq}^{(3)}$ is important as it indicates that
the down-quark channel probes the
orthogonal combination of $C_{lq}^{(1)}$ and $C_{lq}^{(3)}$ compared to the
up-quark channel.

\subsection{Flat directions in Drell-Yan}

The fact that seven Wilson coefficients contribute to the SMEFT
correction but fewer kinematic combinations appear in the matrix
elements implies that only
certain combinations of Wilson coefficients can be probed with
Drell-Yan measurements, a point already made in previous
work~\cite{Alte:2018xgc}.   We can identify the following
features from the above formulae.
\begin{itemize}

  \item The deviations from dimension-6 operators in the SMEFT are
    expected to be largest at high invariant mass, when $\mhs \gg
    M_Z^2$.  When we make this approximation in the denominator of the
    $Z-SMEFT$ interference in Eq.~(\ref{eq:DYMEs}), we find that only two
    combinations of Wilson coefficients proportional to $\mht^2$ and
    $\mhu^2$ respectively contribute.  In the up-quark channel the
    following combinations appear:
    \begin{eqnarray}
-\frac{8 \pi \alpha Q_u }{3}\left[(C_{lu}+C_{qe})\right]+\frac{2g_Z^2
                                }{3}\left[g_R^ug_L^eC_{lu}+g_R^eg_L^uC_{qe}\right],
      \nonumber \\
 -\frac{8 \pi \alpha Q_u }{3}\left[(C_{eu}+C_{lq}^{(1)}-C_{lq}^{(3)})\right]+\frac{2g_Z^2
                                }{3}\left[g_R^ug_R^eC_{eu}+g_L^ug_L^eC_{lq}^{(1)}-g_L^ug_L^eC_{lq}^{(3)}\right].     
    \end{eqnarray}  
    In the down-quark channel similar combinations with the
    replacements of Eq.~(\ref{eq:uptodown}) appear.  In the high-energy limit only these combinations can be probed.

\item In principle these combinations can be separately probed by 
  measurements dependent on the lepton kinematics.  We note that
  measurements of the di-lepton system such as the invariant mass or rapidity do not allow
  these coefficient combinations to be separately determined.  These
  distributions are obtained by integrating inclusively over
  $c_{\theta}$, and in the high-energy limit depend on only a single
  combination of Wilson coefficients.  However, another limitation
  becomes apparent if we express the differential cross section in
  terms of the CM-frame angle $c_{\theta}$.  In the $\mhs \gg M_Z^2$
  limit the SMEFT correction to the cross section takes the form
  \begin{equation}
    A(g_i,C_i) (1+c_{\theta}^2) +B(g_i,C_i) c_{\theta},
  \end{equation}
  where $g_i$ and $C_i$ denote the SM couplings and dimension-6 Wilson
  coefficients respectively.  $A(g_i,C_i)$ is the same combination of couplings that
  appears in the di-lepton invariant mass and rapidity distributions,
  while $B(g_i,C_i)$ is a different combination.  In order to probe $B$ an
  experimental measurement must integrate over an asymmetric range of
  $c_{\theta}$, otherwise the $B$ term will integrate to zero.  Existing
  high-mass differential Drell-Yan measurements that go beyond the
  di-lepton invariant mass and rapidity distributions, such as
  Ref.~\cite{Aad:2016zzw}, focus on quantities such as $|\Delta
  \eta_{ll}|$, the absolute value of the pseudorapidity distribution
  between leptons.  We can express this variable in terms of the
  CM-frame scattering angle as
  \begin{equation}
 |\Delta
  \eta_{ll}| = 2 \, |{\rm arctanh}(c_{\theta})|.
 \end{equation}   
  Since this variable is symmetric under $c_{\theta}
  \to -c_{\theta}$ the $B$ term vanishes.  Other
  measurements of quantities such as the forward-backward asymmetry
  that could distinguish the $B$ term focus primarily on the $Z$-pole
  region or only slightly above
  it~\cite{Khachatryan:2016yte,Aaboud:2017ffb,Sirunyan:2018swq}.  A
  similar point was made in Ref.~\cite{Alte:2018xgc}.  At the $Z$-pole
  all terms except for $ZZ$ interference
 are suppressed by a factor $\Gamma_Z/M_Z$ and are negligible. The
 only sensitivity to $B$ comes from acceptance cuts on the leptons
 which have a small effect on the measured cross section.
    
\end{itemize}

We conclude that the existing Drell-Yan measurements at the LHC can
probe only a limited combination of SMEFT Wilson coefficients.  This
occurs both because of the limited kinematic information available in
the unpolarized Drell-Yan cross section, and also because of the specific measurements performed.
To illustrate this discussion numerically we will consider four
representative combinations
of non-zero Wilson coefficients.
\begin{enumerate}

\item \emph{Case 1}: $C_{eu}, C_{ed}, C_{lq}^{(1)} \ne 0$: these coefficients contribute to
  the $\mht^2$ term in Drell-Yan and can therefore only be
  distinguished by an invariant mass measurement.  They can be separated
  in DIS
  by choosing different electron polarizations according to Eq.~(\ref{eq:DISSMEFT}).

\item  \emph{Case 2}: $C_{qe}, C_{eu},C_{ed} \ne 0$: these are proportional to
  $\mht^2$ and $\mhu^2$ and can therefore in principle be distinguished in
  Drell-Yan, but not with exisiting high-mass LHC measurements.  They can be
  separated by a combination of polarization and differential
  measurements in DIS.

\item \emph{Case 3}: $C_{qe}, C_{lq}^{(1)} \ne 0$: in this case separate
  flat directions appear for the up-quark and down-quark channels that
  cannot be simultaneously satisfied.  We will study this case as a 
  contrast to Cases~1 and~2 in order to determine how much better the
  relevant Wilson coefficients can be probed.  

\item \emph{Case 4}: $C_{lq}^{(1)}, C_{lq}^{(3)} \ne 0$: this is similar to
  Case~3 in that flat directions appear separately in the up-quark and
  down-quark channels.  We study this case to determine how well these
  coefficients can be determined in DIS, where the charged-current
  channel allows a separate measurement of $C_{lq}^{(3)}$.  
  
\end{enumerate}

\section{Fits to Drell-Yan and DIS data} \label{sec:fits}

In order to compare the sensitivities of the EIC and the LHC to
four-fermion Wilson coefficients in the SMEFT, and in particular to study the
ability of the EIC to break the degeneracies present with only Drell-Yan
measurements, we consider fits to the data for the four scenarios defined
above.  

For the Drell-Yan process we consider the data set of Ref.~\cite{Aad:2016zzw}, which
measures the following differential cross sections for invariant
masses up to $1.5$ TeV:
\begin{equation}
  \frac{d\sigma}{d m_{ll}}, \frac{d^2\sigma}{d m_{ll} \, dY_{ll}},  \frac{d^2\sigma}{d m_{ll} \, d|\eta_{ll}|}.
\end{equation}  
We choose this set because it goes to high invariant masses and it
measures the $|\eta_{ll}|$ distribution, allowing us to illustrate our
point above that distributions symmetric under $c_{\theta} \to - c_{\theta}$ offer no discriminatory
power beyond the inclusive invariant mass distribution.  The measurement of $d\sigma/d m_{ll}$ in Ref.~\cite{Aad:2016zzw}
contains twelve bins of invariant mass as compared to five bins of
invariant mass for the two double-differential distributions.  Since
the invariant mass provides the most discriminatory power between
Wilson coefficients we use $d\sigma/d m_{ll}$ in our fits.  We restrict the invariant mass
range to $m_{ll} <700$ GeV in order to have a consistent EFT
expansion for UV scales of $\Lambda \sim 1 \, {\rm TeV}$.  When
performing our fit we use the full experimental correlation matrix
given in Ref.~\cite{Aad:2016zzw}.\footnote{We thank F.~Ellinghaus for
  assistance in understanding the experimental results.}\\

We investigate the same combinations of Wilson coefficients for DIS
observables and contrast the projected EIC bounds with the ones
derived from the Drell-Yan data.  We study the DIS cross sections in
nine separate bins in $(x,Q^2)$ space and assume that the projected
$10\,\textrm{fb}^{-1}$ of collected data is distributed evenly between
the four possible polarized observables: unpolarized and polarized
protons with either choice of electron polarization.  We assume 70\%
polarization for both proton and electron beams. The binning is chosen
so that the statistical error is of the same order as the expected
systematic error.  To obtain the expected cross sections at the EIC we simply evaluate the formulae in
Section~\ref{sec:dis}.  Assuming uncorrelated errors for simplicity we can define a $\chi^2$ statistic according to
\begin{align}
\chi^2 = \sum_{\substack{\lambda_e = \pm 0.7 \\ P=\textrm{pol}/\textrm{unpol}}}{\sum_{i,j}{\left(\frac{\sigma_\textrm{SMEFT}^{\lambda_e, P}(x_i, Q_j)}{\Delta \sigma^{\lambda_e, P}(x_i, Q_j)}\right)^2}}.
\end{align}
The outer sum accounts for the different polarized observables
while the inner sum runs over all bins in $(x,Q^2)$ space. The numerator
denotes the SMEFT-induced deviation in the cross section under
consideration.  For the fits
involving $C_{lq}^{(3)}$ we also include the charged current
observables.  The error $\Delta \sigma^{\lambda_e, P}$ for each of the
observables consists of systematic and statistical errors that we add in
quadrature. We assume the systematic error to be $1\%$ in each bin,
consistent with assumptions in the literature~\cite{Aschenauer:2013iia}.  The
statistical error scales with the collected data.  We assume
$2.5\,\textrm{fb}^{-1}$ to be collected for every observable.  To study
which of the parameter choices impact our fit most strongly we also
present auxiliary fits where we study the effects of increasing the
systematic error, increasing the luminosity, and removing beam
polarizations.  A potential third source of error comes from the 
uncertainties of the PDFs, as discussed earlier. We
choose to omit the PDF errors from our projection since they may ultimately
need to be determined in a simultaneous fit of PDFs and SMEFT
coefficients, as discussed in Ref.~\cite{Carrazza:2019sec}.  They are omitted in our
analysis of LHC data as well for consistency.

\subsection{Case 1}

We begin by studying the behavior of the Drell-Yan cross section for
Case 1 with $C_{eu}$, $C_{ed}$ and $C_{lq}^{(1)}$ non-zero.  All coefficients contribute to the $\mhu^2$ term in the matrix
element, and therefore only the invariant mass distribution can
discriminate between them.  By studying the formulae in
Eq.~(\ref{eq:DYMEs}) we see that the SMEFT correction to the up-quark
channel of the Drell-Yan cross section vanishes for the
following combination of Wilson coefficients in the high
invariant-mass limit:
\begin{equation}
C_{lq}^{(1)} = -C_{eu}\frac{Q_u e^2-g_Z^2 g_R^ug_R^e}{Q_u e^2-g_Z^2 g_L^e g_L^u}
\approx -0.69 C_{eu}.
\label{eq:case1cond1}
\end{equation}
In the down-quark channel the correction vanishes for the combination 
\begin{equation}
C_{lq}^{(1)} = -C_{ed}\frac{Q_d e^2-g_Z^2 g_R^dg_R^e}{Q_d e^2-g_Z^2 g_L^e
  g_L^d} \approx -0.42 C_{ed}.
\label{eq:case1cond2}
\end{equation}
To simplify our analysis of this case we will assume the relation
\begin{equation}
C_{ed} = C_{eu} \frac{Q_u e^2-g_Z^2 g_R^ug_R^e}{Q_u e^2-g_Z^2 g_L^e
  g_L^u} \frac {Q_d e^2-g_Z^2 g_L^eg_L^d}{Q_d e^2-g_Z^2 g_R^dg_R^e},
\equiv C_{ed}^{(1)},
\label{eq:Ced1}
\end{equation}  
which allows both Eqs.~(\ref{eq:case1cond1}) and~(\ref{eq:case1cond2})
to be satisfied. We then allow $C_{lq}^{(1)}$ and $C_{eu}$ to vary.  This choice allows us to
more easily visualize the results of our fits in a two-dimensional
space.  For values of $C_{ed}$ and
$C_{eu}$ that are close to but do not exactly satisfy the relation in Eq.~(\ref{eq:Ced1}) the vanishing of the
SMEFT-induced correction in the high invariant-mass limit will be
approximate.  

\begin{figure}[h!]
  \centering
    \includegraphics[width=4.0in]{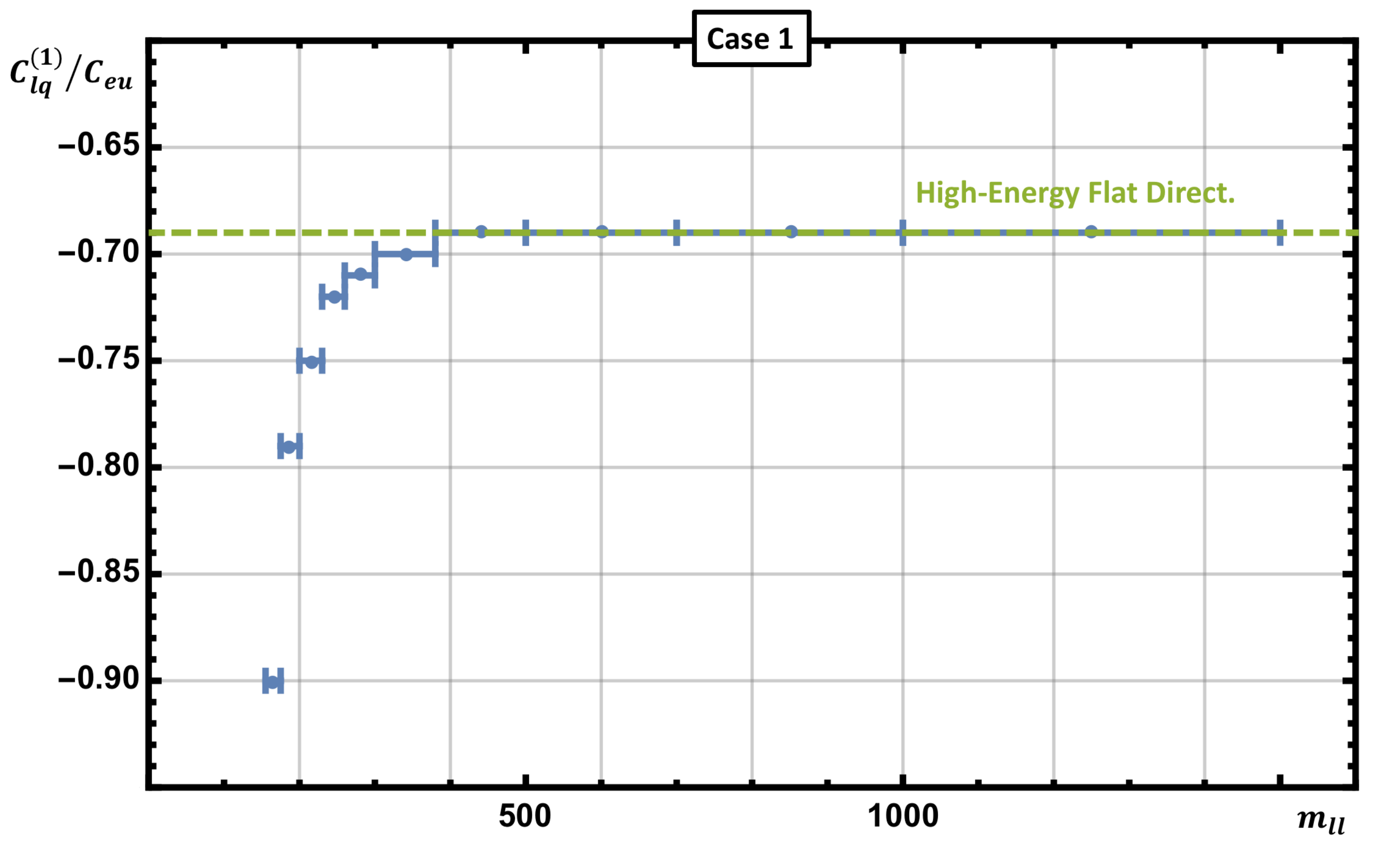}
    \caption{Value of the ratio $C_{lq}^{(1)}/C_{eu}$ for Case 1 for which the SMEFT
      correction to $d\sigma/d m_{ll}$ vanishes as a function of
      the invariant mass bins considered in Ref.~\cite{Aad:2016zzw}.
      This is compared to the predicted value in the $\mhs \gg M_Z^2$
      limit.  The horizontal bars indicate the width of the experimental
      mass bins.}
  \label{fig:zerocross}
\end{figure}

As discussed in the previous section the flat direction for Drell-Yan becomes exact
only when $\mhs \gg M_Z^2$.  In order to check how quickly this limit
is approached we plot in Fig.~\ref{fig:zerocross} the value
of the ratio $C_{lq}^{(1)}/C_{eu}$ for which the SMEFT-induced deviation
vanishes as a function of the invariant mass bin, compared to the
$\mhs \gg M_Z^2$ prediction.   We have assumed $C_{eu}=1/({\rm
  TeV})^2$ when making this plot.  The actual zero crossing approaches the
predicted value quickly as a function of the invariant mass.  This suggests
that this measurement will not strongly probe deviations along this
flat direction, as the high-energy limit where the dimension-6
operators become important coincides with the region where the flat
direction relation is satisfied.

\begin{figure}[h!]
  \centering
    \includegraphics[width=4.0in]{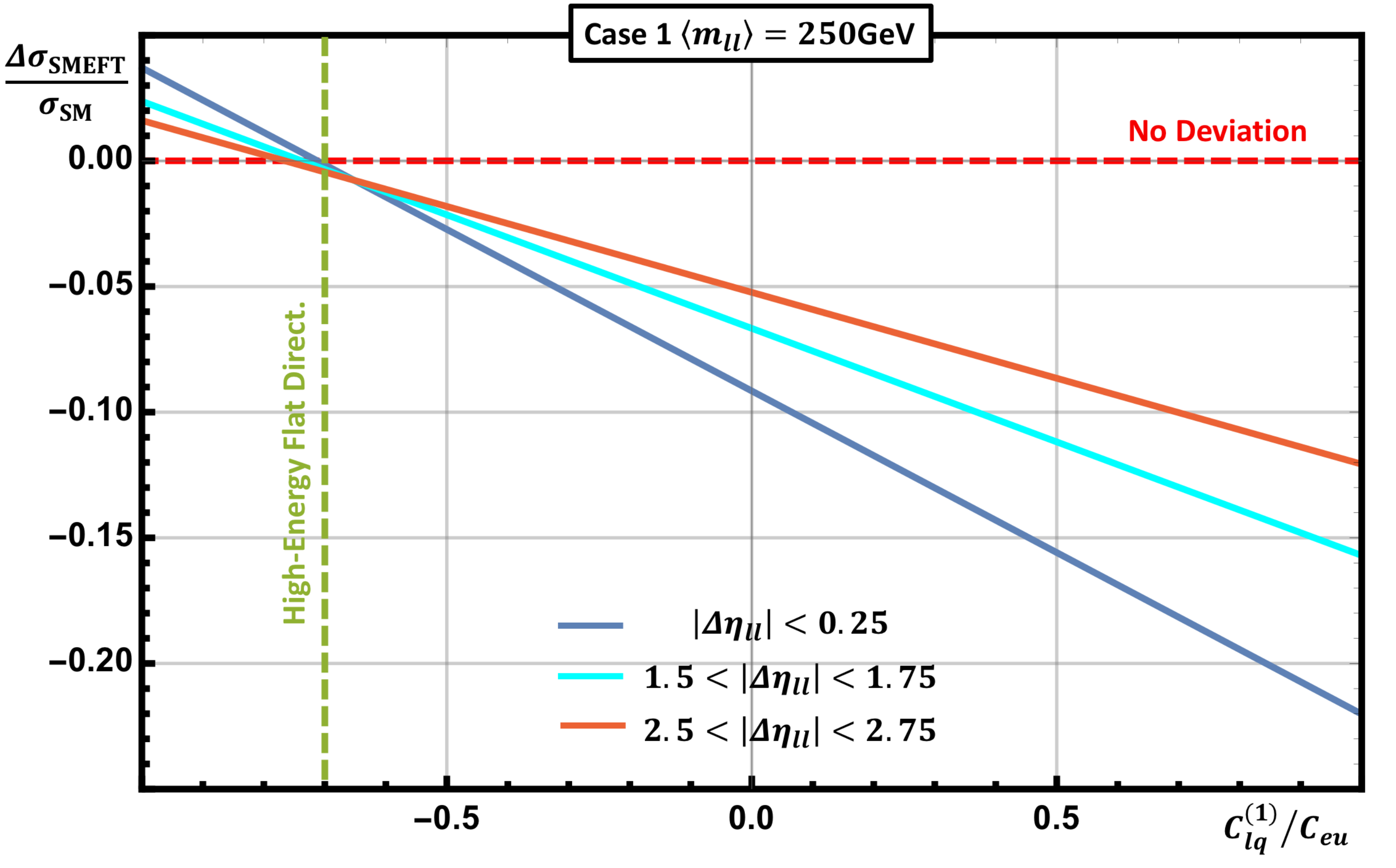}
    \caption{Deviation from the SM as a function of the ratio
      $C_{lq}^{(1)}/C_{eu}$ for the choice $C_{eu}=1/{\rm TeV}^2$ for three
    different $|\Delta \eta_{ll}|$ bins from Ref.~\cite{Aad:2016zzw}.
  The ratio for which no deviation is predicted is also shown.}
  \label{fig:etadev}
\end{figure}

To demonstrate that no additional information is
obtained from the $|\Delta \eta_{ll}|$ distribution as argued in the
previous section, we show in
Fig.~\ref{fig:etadev} the SMEFT-induced deviation for this
distribution as a function of the ratio $C_{lq}^{(1)}/C_{eu}$ for the mass
bin $M_{ll}=[200,300]$ GeV and several choices of $|\Delta \eta_{ll}|$
bins from Ref.~\cite{Aad:2016zzw}.  The deviation vanishes near the predicted
ratio for all choices of $|\Delta \eta_{ll}|$ bins in the experimental
analysis.  Measuring this distribution therefore does not resolve the
flat direction in $C_{lq}^{(1)}$ and $C_{eu}$.

\begin{figure}[h!]
  \centering
    \includegraphics[width=3.0in]{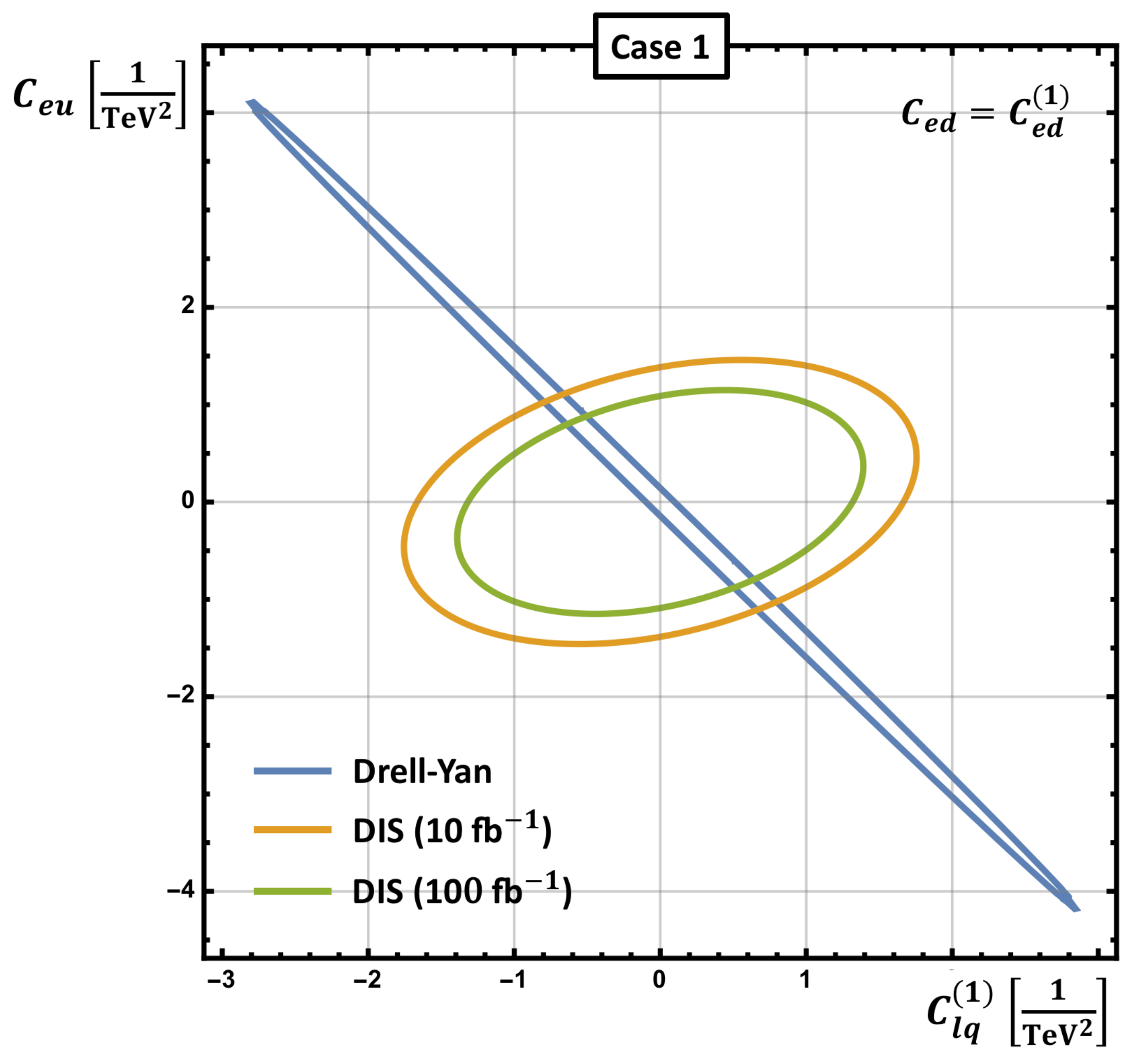}
    \caption{68\% confidence level ellipse in the $C_{lq}^{(1)}$ versus
      $C_{eu}$ space for Case 1.  $C_{ed}$ has been set to the value
      indicated in the text. We contrast the confidence levels derived from the Drell-Yan data with the projected regions for a $10\,\textrm{fb}^{-1}$ and $100\,\textrm{fb}^{-1}$ EIC.}
  \label{fig:case1chi2}
\end{figure}

We now perform separate $\chi^2$ fits to the LHC Drell-Yan and anticipated EIC
data, fixing $C_{ed}$ as discussed above and allowing $C_{lq}^{(1)}$ and
$C_{eu}$ to vary.  The 68\% confidence level (CL) allowed regions are
shown in Fig.~\ref{fig:case1chi2}.  In order to more directly compare the
sensitivities of the two experiments, which is our major goal in this
manuscript, we shift the best-fit values of the Wilson coefficients at
the LHC to the origin.
As we can see in Fig.~\ref{fig:case1chi2} the Drell-Yan data is only
able to constrain the absolute values of $C_{lq}^{(1)}$ and $C_{eu}$
to be smaller than $3.0$ and $4.0$ respectively, in units of $1/ \textrm{TeV}^2$. The projected EIC bounds are more
stringent, $1.5$ and $1.0$ respectively. Increasing the integrated
luminosity
from $10\,\textrm{fb}^{-1}$ to $100\,\textrm{fb}^{-1}$ moderately tightens
the expected bounds.  The plot illustrates that the
two Wilson coefficients are highly correlated in the case of Drell-Yan
observables, as evident from the tight but elongated ellipse. With DIS
data the ellipses are less correlated.  The approximate flat direction is broken through the
interplay of different polarized observables.

\begin{figure}[h!]
  \centering
    \includegraphics[width=3.0in]{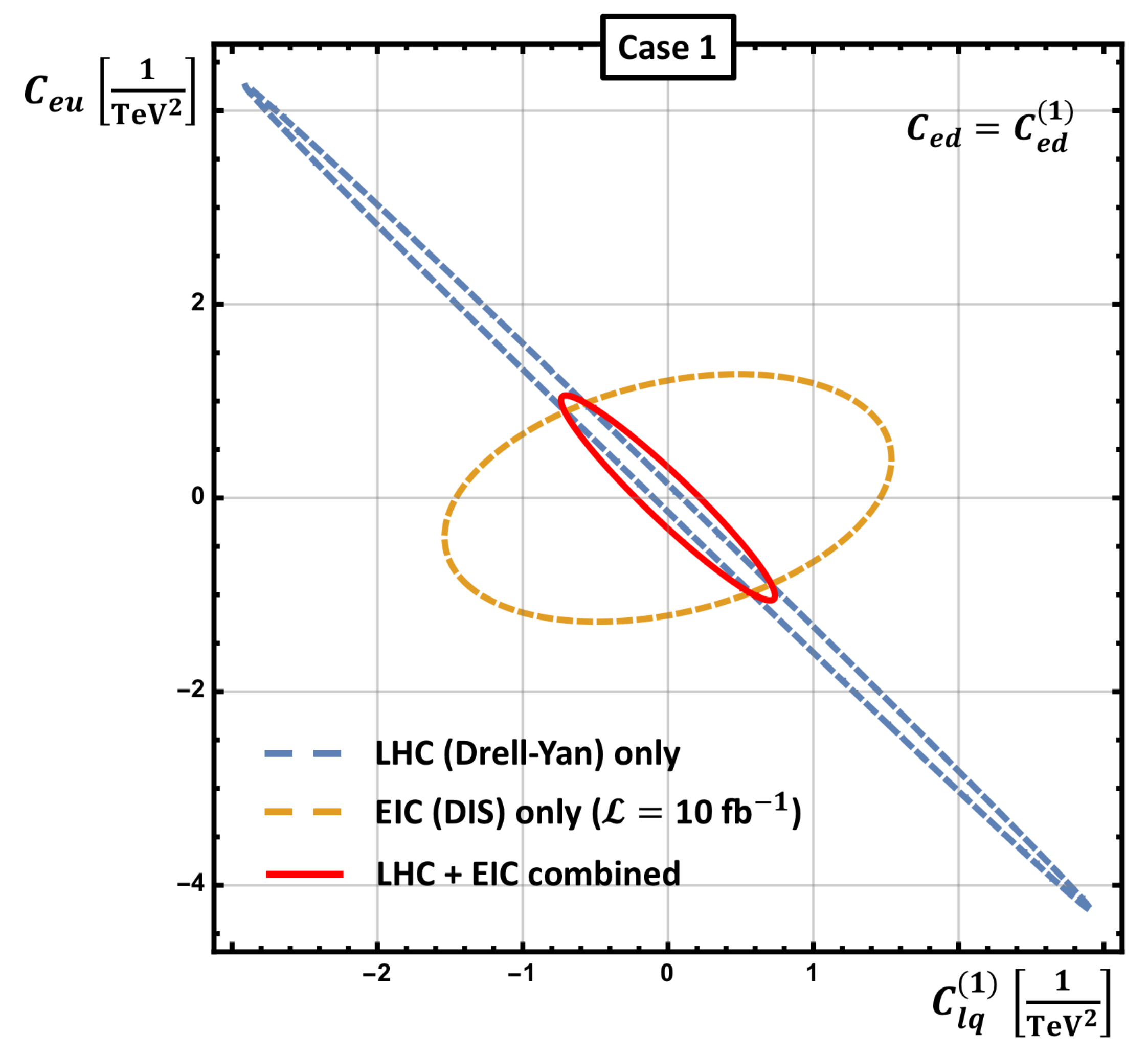}
    \caption{68\% confidence level ellipse in the $C_{lq}^{(1)}$ versus
      $C_{eu}$ space for Case 1 with only LHC data, only EIC data, and
      after combining both experiments.}
  \label{fig:case1combined}
\end{figure}

Ultimately we wish to combine the results from both experiments to
provide the strongest probes of the Wilson coefficients.  This would approximately constrain the possible parameter space
to the overlap of the two respective ellipses.  This is indeed what we
find in Fig.~\ref{fig:case1combined}, where a combined fit to both
the EIC and LHC data sets is compared to each experiment alone.  Each
Wilson coefficient is separately constrained to have a magnitude below
one.  The allowed parameter values along the flat direction poorly
probed by the LHC are reduced by more than a factor of three in this
combined fit.

\subsection{Case 2}

\begin{figure}[h!]
  \centering
    \includegraphics[width=4.0in]{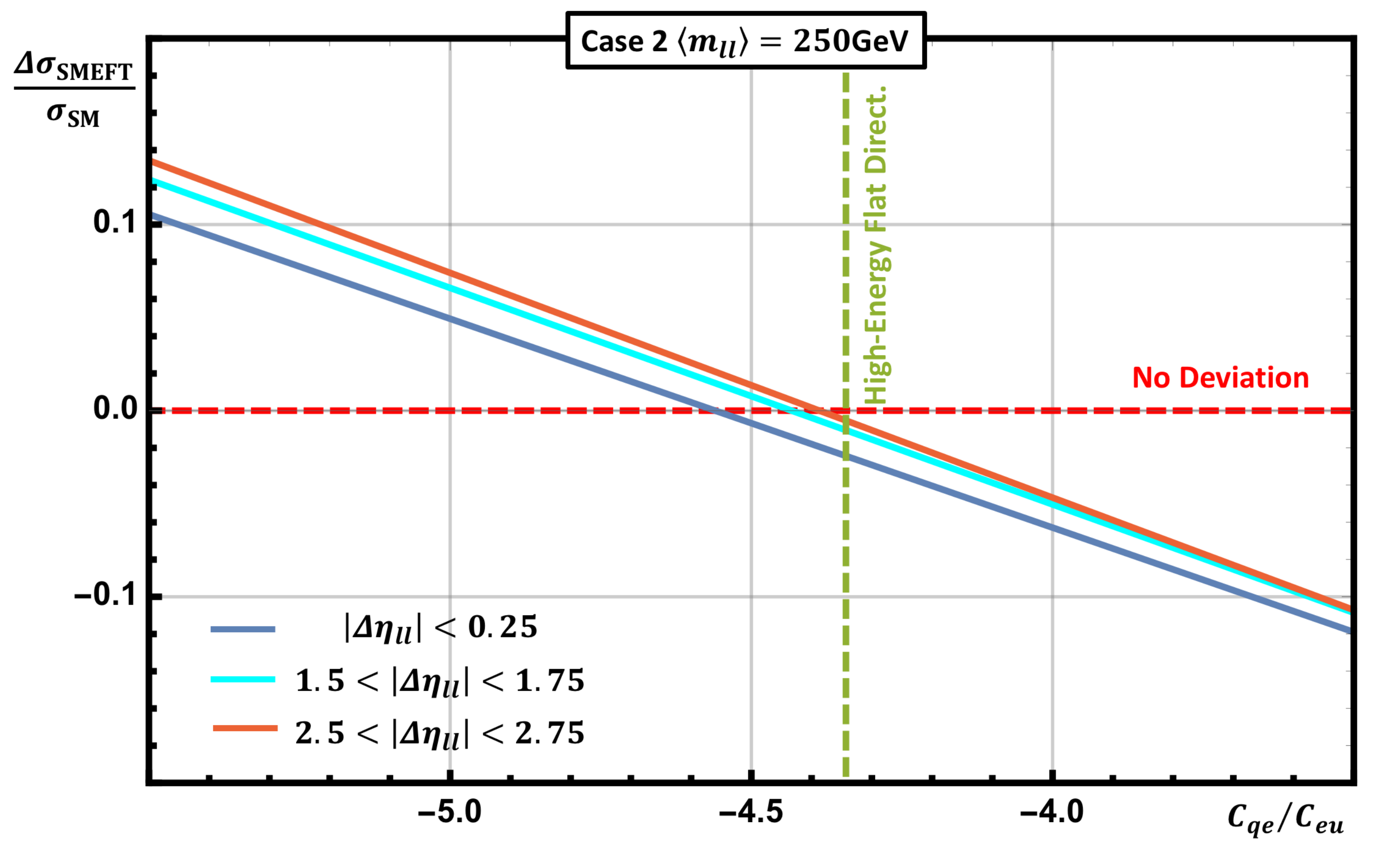}
    \caption{Deviation from the SM as a function of the ratio
      $C_{qe}/C_{eu}$ for the choice $C_{eu}=1/{\rm TeV}^2$ for three
    different $|\Delta \eta_{ll}|$ bins from Ref.~\cite{Aad:2016zzw}.
  The ratio for which no deviation is predicted is also shown.}
  \label{fig:etadevb}
\end{figure}

We next consider the case when $C_{eu}$, $C_{ed}$ and $C_{qe}$ are
non-zero.  Since the $C_{qe}$ dependence of the Drell-Yan matrix element occurs
in the $\mht^2$ term while the other Wilson coefficients contribute to
the $\mhu^2$ terms these coefficients are in principle
distinguishable.  However, as argued above the nature of the studied
experimental measurement at the LHC cannot distinguish between these
coefficients.  To demonstrate this we show in Fig.~\ref{fig:etadevb} the
SMEFT-induced deviation for the $|\Delta \eta_{ll}|$
distribution as a function of the ratio $C_{qe}/C_{eu}$ for the mass
bin $M_{ll}=[200,300]$ GeV.  By integrating over $c_{\theta}$ we can
find the predicted high-energy flat direction:
\begin{eqnarray}
C_{qe} &=& -C_{eu}\frac{Q_u e^2-g_Z^2 g_L^ug_R^e}{Q_u e^2-g_Z^2 g_R^e
           g_R^u} \approx -0.23 C_{eu},
           \nonumber \\
C_{qe} &=& -C_{ed} \frac{Q_d e^2-g_Z^2 g_L^dg_R^e}{Q_d e^2-g_Z^2 g_R^e
           g_R^d} \approx 0.54 C_{ed}.
\end{eqnarray}  
As before we set $C_{ed} = C_{ed}^{(2)}$ with
\begin{equation}
 C_{ed}^{(2)} = C_{eu} \frac{Q_u e^2-g_Z^2 g_L^ug_R^e}{Q_u e^2-g_Z^2 g_R^e
           g_R^u}  \frac {Q_d e^2-g_Z^2 g_R^e
           g_R^d}{Q_d e^2-g_Z^2 g_L^dg_R^e}
\end{equation}  
the value required to
simultaneously satisfy both equations above.  We see in
Fig.~\ref{fig:etadevb} that the SMEFT-induced deviation for all bins
vanishes near the predicted value, again demonstrating that this
distribution cannot discriminate Wilson coefficients near the flat direction.

\begin{figure}[h!]
  \centering
    \includegraphics[width=3.0in]{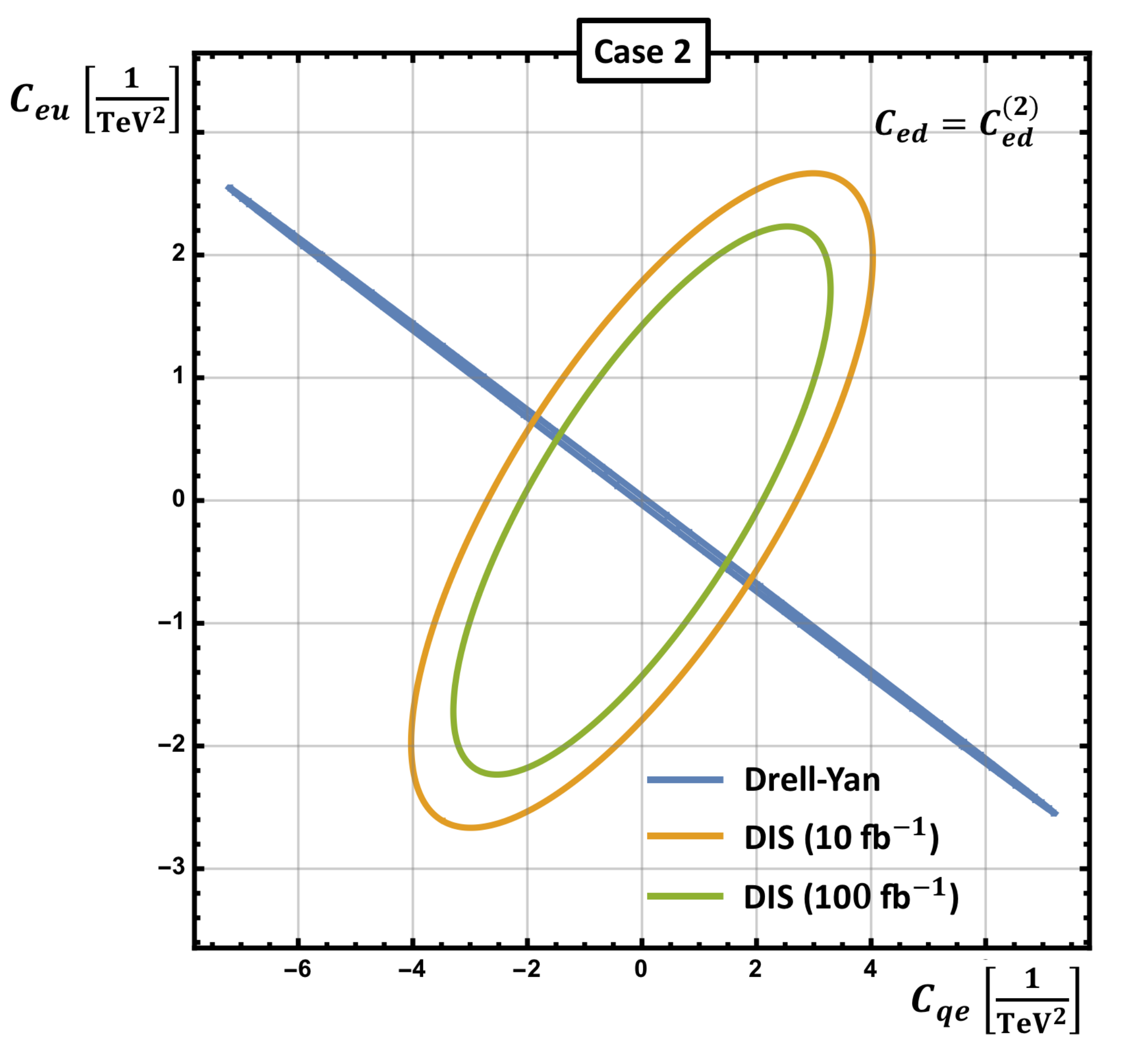}
    \caption{68\% confidence level ellipse in the $C_{qe}$ versus
      $C_{eu}$ space for Case 2.  $C_{ed}$ has been set to the value
      indicated in the text. We contrast the confidence levels derived from the Drell-Yan data with the projected regions for a $10\,\textrm{fb}^{-1}$ and $100\,\textrm{fb}^{-1}$ EIC.}
  \label{fig:case2chi2}
\end{figure}

\begin{figure}[h!]
  \centering
    \includegraphics[width=3.0in]{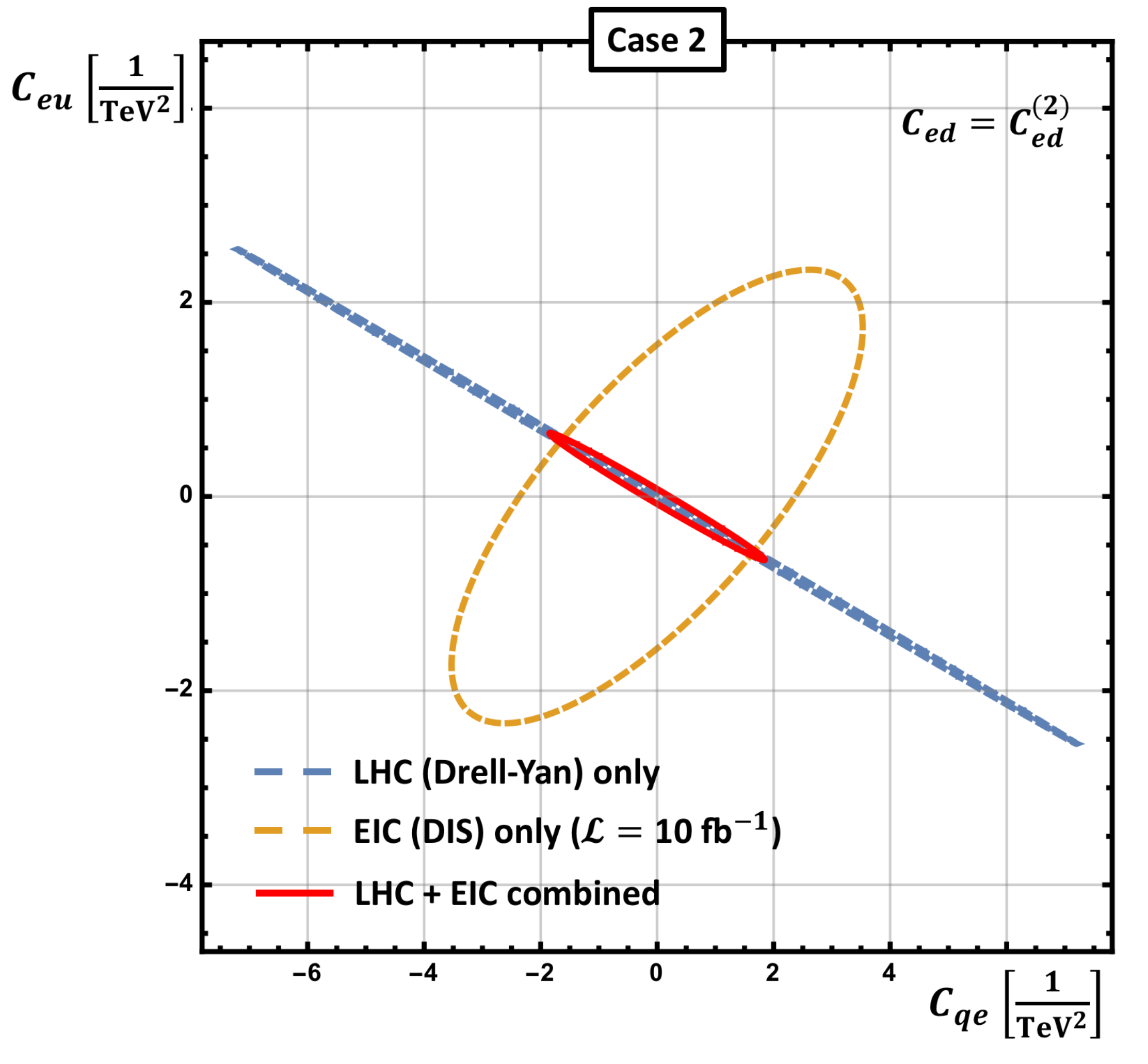}
    \caption{68\% confidence level ellipse in the $C_{qe}$ versus
      $C_{eu}$ space for Case 2 with only LHC data, only EIC data, and
      after combining both experiments.}
  \label{fig:case2combined}
\end{figure}

We perform similar $\chi^2$ fits as done for Case 1 above. The resulting bounds can be seen in Fig.~\ref{fig:case2chi2}.
Similar to Case 1 the Drell-Yan data constrains the absolute values of
$C_{qe}$ and $C_{eu}$ to be smaller than roughly $7$ and $2.5$
respectively in units of $1/\textrm{TeV}^2$. The EIC ellipses are similar in magnitude with a
projected constraint for $C_{qe}$ between about $-4$ and $4$ and $-2.5$
and $2.5$ for $C_{eu}$. Once again, since there is a flat
direction present in the Drell-Yan expressions the corresponding
ellipse is highly correlated, which is not the case at the EIC.
Combining both experiments would allow us to constrain both parameters
to roughly unity.  We show this in Fig.~\ref{fig:case2combined} with a combined fit to both
the EIC and LHC data sets, compared to each experiment alone.  The
ellipse indicating the allowed region is still elongated along the
direction poorly probed by the Drell-Yan data, but the allowed
parameter values are reduced by nearly a factor of three.

\subsection{Case 3}

\begin{figure}[h!]
  \centering
    \includegraphics[width=3.0in]{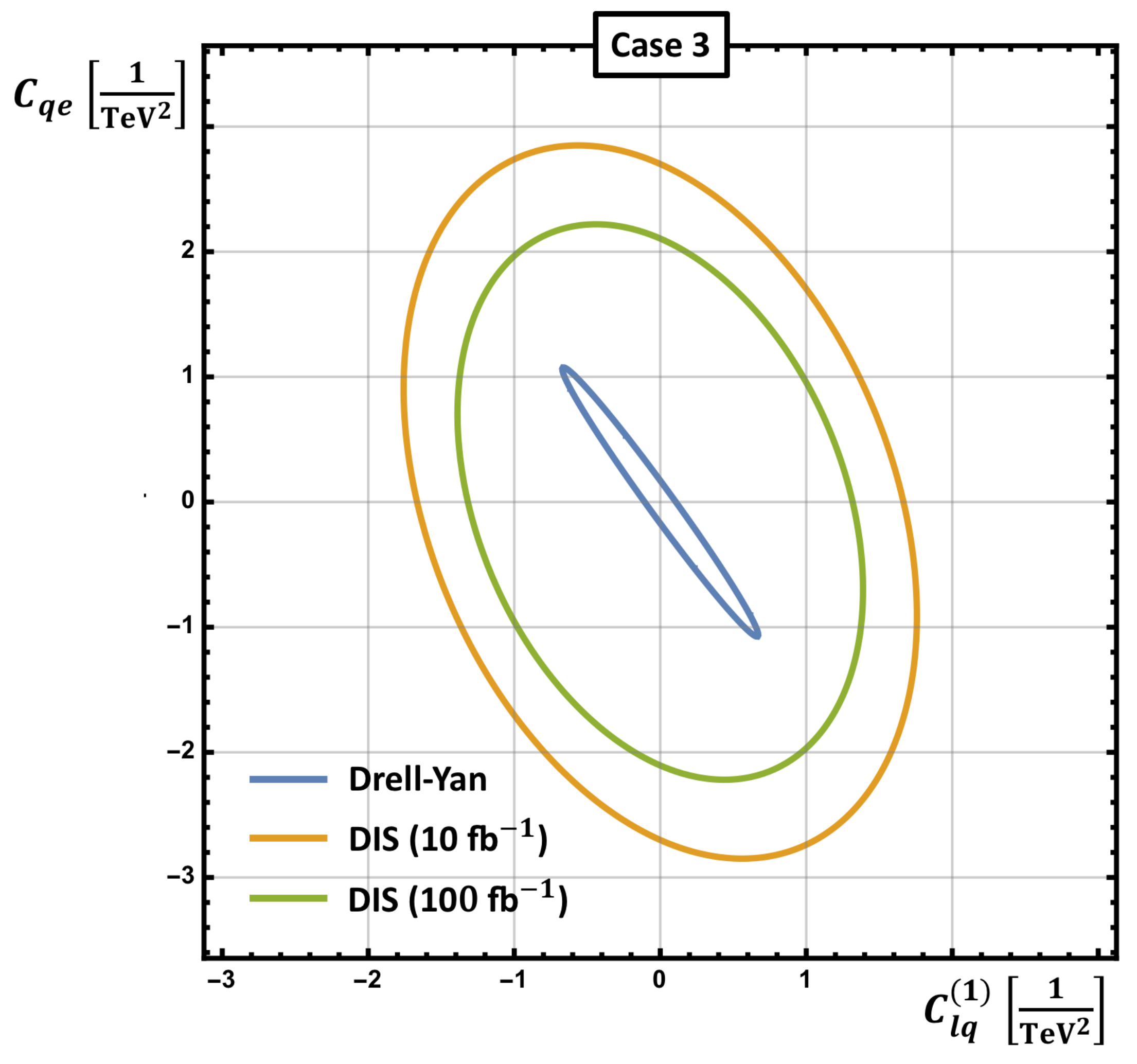}
    \caption{68\% confidence level ellipse in the $C_{qe}$ versus
      $C_{lq}^{(1)}$ space for Case 3. We contrast the confidence levels derived from the Drell-Yan data with the projected regions for a $10\,\textrm{fb}^{-1}$ and $100\,\textrm{fb}^{-1}$ EIC.}
  \label{fig:case3chi2}
\end{figure}

We now consider Case 3, where both $C_{qe}$ and $C_{lq}^{(1)}$ are
non-zero.  In the high-energy limit of the Drell-Yan cross section, flat directions exist separately
in the up-quark and down-quark channels.  Assuming a symmetric
integration over $c_{\theta}$ they can be found to be:
\begin{itemize}

\item Up-quark: 
  \begin{equation}
C_{lq}^{(1)} = -C_{qe}\frac{Q_u e^2-g_Z^2 g_L^ug_R^e}{Q_u e^2-g_Z^2g_L^e g_L^u}
\approx -0.16 C_{qe}.
\end{equation}

\item Down-quark: 
  \begin{equation}
C_{lq}^{(1)} = -C_{qe}\frac{Q_d e^2-g_Z^2 g_L^dg_R^e}{Q_u e^2-g_Z^2g_L^e g_L^d}
\approx 0.22 C_{qe}.
\end{equation}

\end{itemize}  
These two equations for $C_{lq}^{(1)}$ and $C_{qe}$ cannot be
simultaneously satisfied, indicating that Drell-Yan measurements
should be able to better probe this choice of parameters than the two
cases considered previously.  We show the projected bounds in Fig.~\ref{fig:case3chi2}.
The EIC bounds derived for $C_{lq}^{(1)}$ and $C_{qe}$ are similar to
the ones in Case 1 and 2 and constrain the absolute values of the
coefficients to be smaller than about $1.5$ and $2.5$
respectively. There is very little correlation between the
coefficients as evident from the ellipses.  The Drell-Yan bounds are
significantly tighter than the DIS bounds in Case 3. This is
expected since there is no flat direction involving $C_{lq}^{(1)}$ and
$C_{qe}$ in Drell-Yan.  This case illustrates the power of the
Drell-Yan data in the absence of flat directions; the bounds obtained
are nearly an order of magnitude stronger than the Drell-Yan bounds found for Cases~1 and~2.

\subsection{Case 4}

Finally, we consider $C_{lq}^{(1)}$ and $C_{lq}^{(3)}$ non-zero.  The
Drell-Yan up-quark channel depends on the combination $C_{lq}^{(1)}-C_{lq}^{(3)}$ while
the down-quark channel depends on $C_{lq}^{(1)}+C_{lq}^{(3)}$.  In principle
these two Wilson coefficients are distinguishable through the
different kinematics of these two channels.  In DIS we can directly
access $C_{lq}^{(3)}$ through charged-current scattering.  At the LHC
this would be possible if an analysis similar to
Ref.~\cite{Aad:2016zzw} were performed for off-shell $W$-boson
production.  We are not aware of such an analysis.
\begin{figure}[h!]
  \centering
    \includegraphics[width=3.0in]{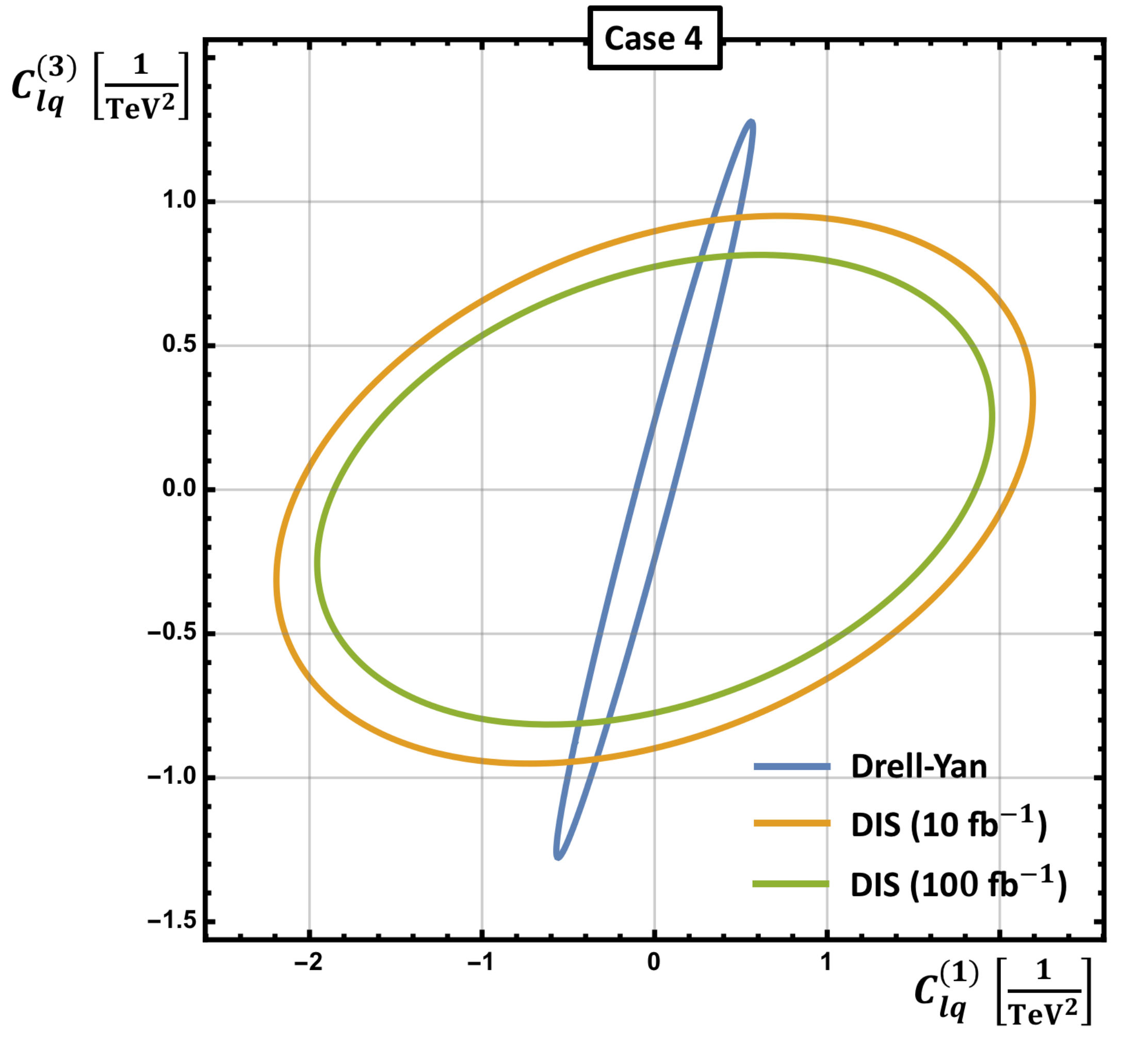}
    \caption{68\% confidence level ellipse in the $C_{lq}^{(3)}$ versus
      $C_{lq}^{(1)}$ space for Case 4. We contrast the confidence levels derived from the Drell-Yan data with the projected regions for a $10\,\textrm{fb}^{-1}$ and $100\,\textrm{fb}^{-1}$ EIC.}
  \label{fig:case4chi2}
\end{figure}
We present the bounds obtained for the LHC and EIC in
Fig.~\ref{fig:case4chi2}.  The bounds obtained from the Drell-Yan data
are stronger than in Cases~1 and~2.  This finding is consistent with the absence of
a flat direction due to the different dependences on $C_{lq}^{(1)}$
and $C_{lq}^{(3)}$.  The constraint on $C_{lq}^{(1)}$ reaches below
$1$.  The bounds can be improved through the inclusion of EIC
charged-current data which are exclusively sensitive to
$C_{lq}^{(3)}$.

\subsection{Effects of parameter choices on EIC fits}

\begin{figure}[h!]
  \centering
    \includegraphics[width=5.0in]{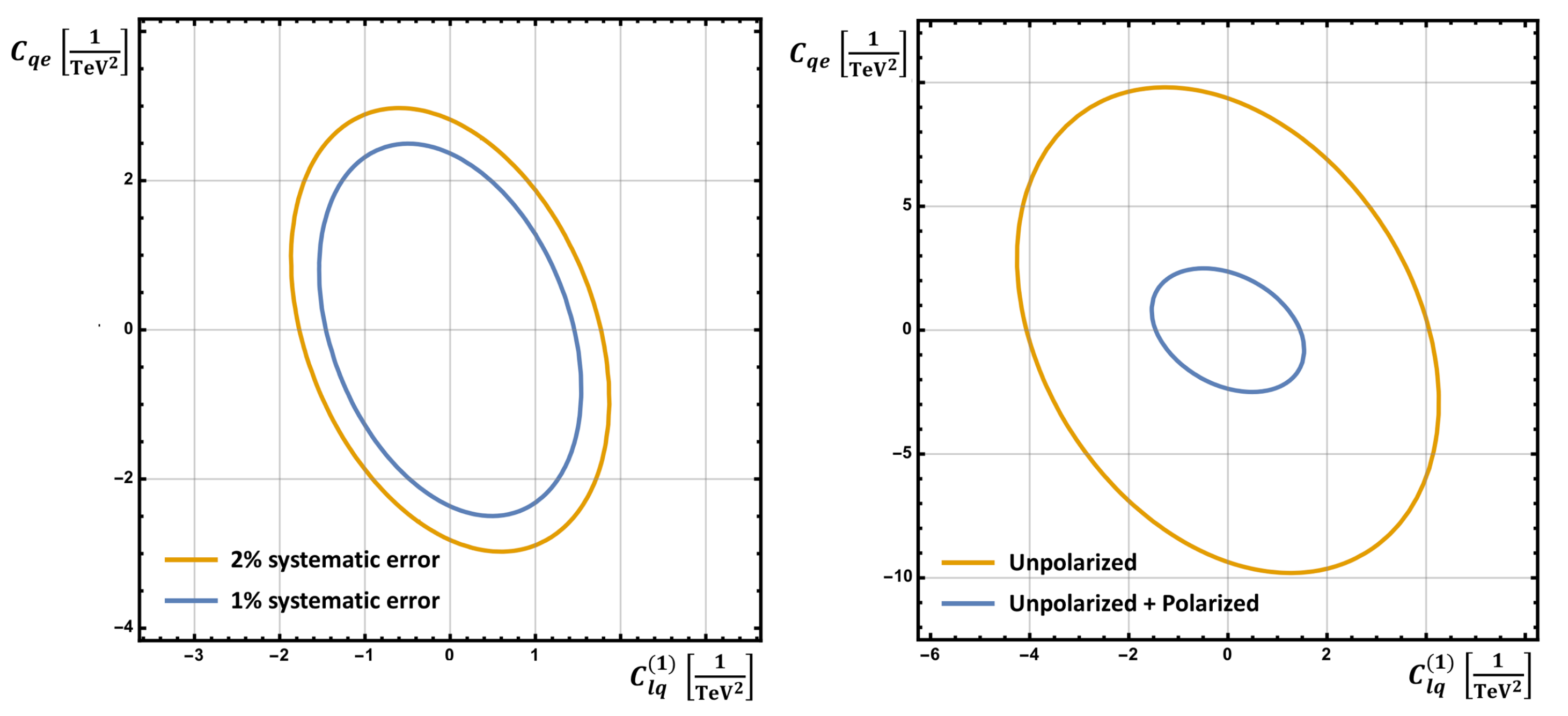}
    \caption{Example $68\%$ CL ellipses for different choices of the
      systematic error (left) and upon removing polarized observables (right).}
  \label{fig:SysPol}
\end{figure}

We study here the impact of EIC systematic error and polarization on
the results obtained above, using Case~3 as a representative example.
Our results are shown in Fig.~\ref{fig:SysPol}.  We see that
increasing the systematic error from 1\% to 2\% has little impact on
the analysis.
  However, it is clear that the ability of the EIC to measure
  polarized observables is crucial in obtaining strong probes of Wilson
  coefficients.  The projected bounds weaken by a factor of four if
  polarized observables are removed from the fit.  We have
  additionally investigated the effect of increasing the polarization
  of the electron beam to $\lambda_e = 0.85$ and $\lambda_e = 0.95$.
  The impact on the bounds and the correlation of the ellipses is
  negligible.
%

\section{Conclusions}  \label{sec:conc}

We have studied in this paper the potential of future EIC measurements
to probe dimension-6 operators in the SMEFT.  The possibility of
measuring polarized cross sections at an EIC provides a powerful handle on
four-fermion operators in the SMEFT.  In particular, the ability to
measure both the unpolarized and
polarized proton cross sections with different electron polarizations
allows the effects of different Wilson coefficients to be disentangled.  This discrimination between dimension-6 effects is not
possible with just Drell-Yan data at the LHC, where only limited
combinations of Wilson coefficients are accessible.  In addition,
the absence of high invariant-mass measurements of quantities such as
a forward-backward asymmetry at the LHC further limits the ability of
the Drell-Yan data to disentangle the various dimension-6 effects.  We
demonstrate these points by example fits to both available LHC data
and projected EIC data in four different scenarios that illustrate the
flat directions present with only Drell-Yan invariant mass data
available.  We show that fits including both LHC and future EIC data
provide much stronger constraints on the Wilson coefficients than fits
to either experiment separately.

\section*{Acknowledgments}
We thank F.~Ellinghaus for helpful comments on the treatment of
correlated errors in the ATLAS high invariant-mass Drell-Yan data.
R.~B. is supported by the DOE contract DE-AC02-06CH11357.  F.~P. and D.~W. are supported
by the DOE grants DE-FG02-91ER40684 and DE-AC02-06CH11357.
The U.S. Government retains for itself, and others acting on its behalf, a paid-up nonexclusive, irrevocable worldwide license in said article to reproduce, prepare derivative works, distribute copies to the public, and perform publicly and display publicly, by or on behalf of the Government.

\end{document}